\documentclass[a4paper,cite]{article}
\usepackage{latexsym}

\usepackage[latin1]{inputenc}

\usepackage{graphicx}
\usepackage{color,pst-plot}
\usepackage{amsmath,amsfonts,amssymb,amstext,amscd,amsthm,dsfont}

\newcommand{\lbl}[1]{\label{eq:#1}}
\newcommand{ \rf}[1]{(\ref{eq:#1})}

\newcommand{\be}{\begin{equation}}
\newcommand{\ee}{\end{equation}}
\newcommand{\bea}{\begin{eqnarray}}
\newcommand{\eea}{\end{eqnarray}}

\newcommand{\noi}{\noindent}
\newcommand{\nn}{\nonumber}
\newcommand{\ra}{\rightarrow}

\newcommand{\cF}{{\cal F}}

\newcommand{\cO}{{\cal O}}

\newcommand{\tr}{\mbox{\rm tr}}

\newcommand{\ksls}{\not \! k}
\newcommand{\psls}{\not \! p}

\newcommand{\pslsh}{\not \! p}

\newcommand{\pslsin}{\not \! p_{1}}
\newcommand{\pslsout}{\not \! p_{2}}

\input epsf
\begin{document}

\title{~\\[2cm]
\bf\Large
Hadronic Light--by--Light Scattering Contribution\\[1mm]
{\Large\bf to the Muon Anomalous Magnetic Moment}}
\author{~\\ Joaquim Prades$^{\,a}$, Eduardo de Rafael$^{\,b}$ and  Arkady Vainshtein$^{\,c}$\\[2mm]
{\small \sl $^{a}$CAPFE and Departamento de F\'isica  Te\'orica y del Cosmos,}\\[-1mm]
{\small \sl Universidad de Granada, Campus de Fuente Nueva, E-18002 Granada, Spain}\\
{\small \sl $^b$Centre de Physique Th\'eorique, CNRS-Luminy Case 907, F-13288 Marseille Cedex 9, France}\\
{\small \sl $^c$William I. Fine Theoretical Physics Institute, University of Minnesota,
Minneapolis, MN 55455, USA}}
\date{}                                           

\maketitle
\thispagestyle{empty}

\vspace{-10.5cm}

\begin{flushright}
UG-FT/242-08\\ 
CAFPE/112-08\\
CPT-P092-2008\\
FTPI-MINN-08/41\\
UMN-TH-2723/08
\end{flushright}

\vspace{7.5cm}

\begin{abstract}
We review the current status of theoretical calculations of the hadronic light-by-light scattering contribution to the muon anomalous magnetic moment.
Different approaches and related issues such as OPE constraints and large breaking of 
chiral symmetry  are discussed. Combining results of different models with educated guesses
on the errors we come to the estimate $$a^{\rm HLbL}=(10.5\pm 2.6)\times 10^{-10}\,.$$
The text  is prepared as a contribution  
 to the {\it Glasgow White Paper on the present status of the Muon Anomalous Magnetic Moment}.
\end{abstract}
 \newpage

\noindent
{{\bf 1. Introduction.}\\[2mm]
From a  theoretical point of view the hadronic light--by--light scattering (HLbL) contribution to the muon magnetic moment is described by the vertex function (see Fig.\,1 below):
\be
\Gamma_{\mu}^{(H)}(p_2 , p_1) =  i e^6\!\! \int\!\!\frac{d^4 k_1}{(2\pi)^4}\!\!\int\!\!\frac{d^4 k_2}{(2\pi)^4}\frac{\Pi_{\mu\nu\rho\sigma}^{(H)}(q,k_1, k_3, k_2)}{k_1^2 k_2^2 k_3^2}\ \gamma^{\nu}(\pslsout+\ksls_{2}-m_{\mu})^{-1}\gamma^{\rho}(\pslsin -\ksls_{1}-m_{\mu})^{-1}\gamma^{\sigma}\,,
\label{vertex}
\ee
where $m_{\mu}$ is the muon mass and $\Pi_{\mu\nu\rho\sigma}^{(H)}(q,k_1, k_3, k_2)$, with $q=p_{2}-p_{1}=-k_1 -k_2 -k_3$, denotes the off--shell photon--photon scattering amplitude induced by hadrons,
\bea
\Pi_{\mu\nu\rho\sigma}^{(H)}(q,k_1, k_3, k_2)	&\!\! =\!\! & 
\!\!\int\!\! d^4x_1\!\!\int\!\! d^4x_2\!\!\int\!\! d^4x_3\  \exp[{-i(k_1\cdot x_1\! +\!k_2\cdot x_2\!+\!k_3\cdot x_3)]} 
\nn \\[1mm]
 &  & \times\langle 0 \vert T\{j_{\mu}(0)\ j_{\nu}(x_1)\ j_{\rho}(x_2)\  j_{\sigma}(x_3)\}\vert 0\rangle\,.
 \eea
Here  $j_{\mu}$ is  the Standard Model electromagnetic current, 
$j_{\mu}(x)=\sum_{q}Q_{q}\bar q(x) \gamma_{\mu}q(x)$, where $Q_{q}$ denotes the electric charge of quark $q$.
\begin{figure}[h]

\begin{center}
\includegraphics[width=0.3\textwidth]{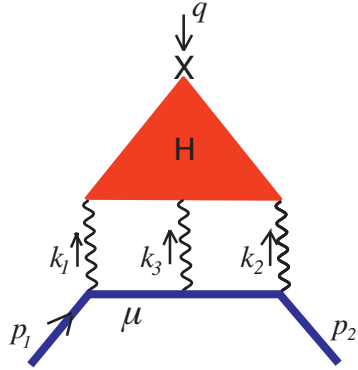}

\vspace*{0.2cm}
\caption{\it\small  Hadronic light--by--light scattering contribution.}\label{figure:llbyl}
\end{center}

\end{figure}
The external photon with momentum $q$ represents the magnetic field. We are interested
in the limit $q\to 0$ where  the current conservation implies that $\Gamma_{\mu}^{(H)}$ is linear in $q$,  
\be
\Gamma_{\mu}^{(H)}=-\frac{a^{\rm HLbL}}{4m_{\mu}}\,[\gamma_{\mu}\,,\gamma_{\nu}]\,q^{\nu}\,.
\ee
The muon anomaly can then be extracted as follows
\bea
\lbl{aH}
a^{\rm HLbL}\!\!\!\!&=&\!\!\!\!\frac{-ie^{6}}{48m_{\mu}}\int\!\frac{d^4 k_1}{(2\pi)^4}\!\!\int\frac{d^4 k_2}{(2\pi)^4}\frac{1}{k_1^2 k_2^2 k_3^2}
\left[\frac{\partial}{\partial q^{\mu}}\Pi_{\lambda\nu\rho\sigma}^{(H)}(q,k_1, k_3, k_2)\right]_{q=0}
 \nn \\[1mm] 
 & &\times \,\tr\left\{
        (\psls + m_{\mu})[\gamma^{\mu},\gamma^{\lambda}](\psls +m_{\mu}) \gamma^{\nu}(\pslsh+\ksls_2-m_{\mu})^{-1}\gamma^{\rho}(\pslsh -\ksls_1-m_{\mu})^{-1}\gamma^{\sigma}\right\}\,.
\eea

Unlike the case of the hadronic vacuum polarization (HVP) contribution, there is no direct experimental input for the hadronic light--by--light scattering (HLbL) so one has to rely
on theoretical approaches. Let us start with the massive quark loop contribution 
which is known analytically,
\be
a^{\rm HLbL}({\rm quark~loop})=\left(\frac{\alpha}{\pi}\right)^{3}\!\!N_{c} Q_{q}^{4}\,\Bigg\{ \underbrace{\left[\frac{3}{2}\,\zeta(3)-\frac{19}{16} \right]}_{0.62}\frac{m_{\mu}^{2}}{m_{q}^{2}}+\cO\left[\frac{m_{\mu}^{4}}{m_{q}^{4}} \log^{2} \!\frac{m_{\mu}^{2}}{m_{q}^{2}}\right]\Bigg\}\,,
\ee
 where $N_{c}$ is the number of colors and $m_{q}\gg m_{\mu}$ is implied. It gives a reliable result for the heavy quarks $c\,,b\,,t$ with $m_{q}\gg \Lambda_{\rm QCD}$.  Numerically, however, heavy quarks do not contribute much. For the $c$ quark, with $m_{c}\approx 1.5~{\rm GeV}$, 
 \be\lbl{charm}
 a^{\rm HLbL}({\rm c})=0.23 \times 10^{-10}\,.
 \ee
 To get a very rough estimate for the light quarks $u,d,s$ let us use a constituent mass of 300 MeV for $m_{q}$\,. This gives 
 $a^{\rm HLbL}(u,d,s)= 6.4\times 10^{-10}$.
QCD tells us that the quark loop should be accurate in describing large virtual momenta, $k_{i}\gg \Lambda_{\rm QCD}$, i.e. short--distances. What is certainly missing in this constituent quark loop estimate, however,  is the low--momenta piece dominated by a neutral pion--exchange in the light--by--light scattering. Adding up this contribution, discussed in more detail below, approximately doubles the estimate to
$a^{\rm HLbL}\approx 12 \times 10^{-10}$. While the ballpark of the effect
is given by this rough estimate, a more refined analysis is needed to get its magnitude and evaluate the accuracy. Details and comparison of different contributions will be discussed below, but it is already interesting to point out that all existing calculations fall into a range:
\be
a^{\rm HLbL}=(11\pm 4) \times 10^{-10}\,,
\ee
compatible with this rough estimate.
The dispersion of the $a^{\rm HLbL}$ results in the literature is not too bad when compared with the present experimental accuracy of $6.3\times 10^{-10}$. However the proposed new $g_{\mu}\!-\!2$ experiment sets a  goal of $1.4\times 10^{-10}$ for the error, which calls for a considerable improvement in the theoretical calculations as well. We believe  that theory 
is up to this  challenge; a further use of theoretical and experimental constraints could result in reaching such accuracy soon enough.

The history of the evaluation of the hadronic light--by--light scattering contribution is a long one which can be found in the successive review articles on the subject. In fact, but for the sign error in  the neutral pion exchange discovered in 2002~\cite{KN02,KNPdeR02}, the theoretical predictions for $a^{\rm HLbL}$ have been relatively stable over more than ten years.  
 
Here we are interested in highlighting the  generic properties of QCD relevant to the evaluation of Eq.~\rf{aH}, as well as  their connection with the most recent model dependent estimates which have been made so far.  

\vspace*{0.6cm}
\noi
{\bf 2. QCD in the Large $\mathbf{N_c}$ and Chiral Limits }\\[2mm]
For the light quark components in the electromagnetic current ($q=u\,, d\,, s$) the integration of the light--by--light scattering over virtual momenta in Eq.\,\rf{aH} is convergent at characteristic hadronic scales.
We choose the mass of the $\rho$ meson $m_{\rho}$ to represent that scale.  Of course, hadronic 
physics at such momenta is non--perturbative and the first question to address is what theoretical parameters can be used to define an expansion. Two possibilities are: the large number of colors, $1/N_{c}\ll 1$, and the smallness of the chiral symmetry breaking, $m_{\pi}^{2}/m_{\rho}^{2}\ll 1$. Their relevance  can be seen from the expansion of $a^{\rm HLbL}$ as a power series in these parameters,
\be
\label{param}
a^{\rm HLbL}\sim \Big(\frac{\alpha}{\pi}\Big)^{3}\frac{m_{\mu}^{2}}{m_{\rho}^{2}}\,\Big[c_{1}\,N_{c}+c_{2}\,\frac{m_{\rho}^{2}}{m_{\pi}^{2}} +c_{3}+{\cal O}(1/N_{c})\Big]\,,
\ee
where $m_{\pi} > m_{\mu}$ is implied. Only the power dependencies are shown;
possible chiral logarithms, $\ln(m_{\rho}/m_{\pi})$, are included into the coefficients $c_{i}$.\\[0.1cm]

\noi
{\sc Terms leading in the large $N_{c}$ limit}\\[2mm]
The first term, linear in $N_{c}$\,,  comes from the one--particle exchange of a meson $M$
in the HLbL amplitude, see Fig.\,2(a). 
\begin{figure}[h]

\begin{center}
\includegraphics[width=0.8\textwidth]{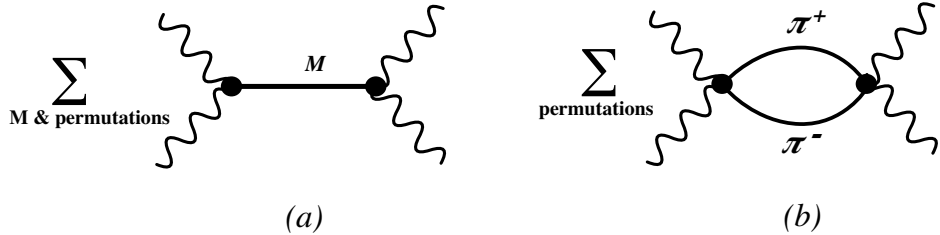}

\caption {\it\small Diagrams for HLbL: (a)  meson exchanges, (b) the charged pion loop, 
the blob denotes the full $\gamma^{*}\gamma^{*} \to \pi^{+}\pi^{-}$ amplitude. \label{lbl}}
\end{center}
\end{figure}
In principle, the meson $M$ is any neutral, C--even meson. In particular this includes 
pseudoscalar mesons
$\pi^{0},\,\eta,\, \eta'$; scalars $f_{0},\, a_{0}$; vectors $\pi_{1}^{0}$; pseudovectors $a_{1}^{0},\,f_{1},\,f_{1}^{*}$;
spin 2 tensor and pseudotensor mesons  $f_{2},\,a_{2},\,\eta_{2},\,\pi_{2}$\,. 

The neutral pion  
exchange is special because of the Goldstone nature of the pion; its mass is much smaller
than the hadronic scale $m_{\rho}$.  In $a^{\rm HLbL}(\pi^{0})$ this leads to an additional enhancement  by two powers of a chiral logarithm \cite{KNPdeR02}, 
\be
\lbl{piefft}
a^{\rm HLbL}(\pi^{0})=\Big(\frac{\alpha}{\pi}\Big)^3\!N_{c}\,\frac{m_{\mu}^{2}N_{c}}{48\pi^2 F_{\pi}^{2}}
\Big[ \ln^2\frac{m_{\rho}}{m_{\pi}}  +
\cO\Big(\ln\frac{m_{\rho}}{m_{\pi}}\Big)+\cO(1)\Big]
\,.
\ee
Here the $\pi^{0}\gamma\gamma$ coupling is fixed by the Adler--Bell--Jackiw anomaly in terms of
the pion decay constant $F_{\pi}\approx 92~{\rm MeV}$. This constant is $\cO\left(\sqrt{N_{c}}\right)$, therefore $N_{c}/F_{\pi}^{2}$  behaves as a constant in the large--$N_c$ limit\,. The mass of the $\rho$ plays the role of an ultraviolet scale
in the integration over $k_{i}$ in Eq.~\rf{aH} while the pion mass provides the infrared scale. 
Of course, the muon mass is also important at low momenta but one can keep the ratio $m_{\mu}/m_{\pi}$ fixed in the chiral limit. 

Equation \rf{piefft} provides the result for $a^{\rm HLbL}$ for the term leading in the $1/N_{c}$ expansion in the chiral limit where the pion mass is much less than the next hadronic scale. In this limit the dominant neutral pion exchange produces the characteristic universal double logarithmic behavior with the {\em exact} 
coefficient given in Eq.~\rf{piefft}. Testing this limit was particularly useful in fixing the sign of the neutral pion exchange.

Although the coefficient of the $\ln^2 (m_{\rho}/m_{\pi})$ term in Eq.\,\rf{piefft} is unambiguous, the coefficient
of the $\ln(m_{\rho}/m_{\pi})$ term depends on low--energy constants which are difficult to extract from experiment~\cite{KNPdeR02,RMW02}  (they require a detailed knowledge of the $\pi^0 \ra e^+ e^-$ decay rate with inclusion of radiative corrections). 
Model dependent estimates of the single logarithmic term as well as the constant term show 
that these terms are not suppressed. It means that we cannot rely on chiral perturbation theory
and have  to adopt a dynamical framework which takes into account explicitly the heavier meson exchanges as well.

Note that the overall sign of the pion exchange, for physical values of the masses, is much less model dependent than the previous chiral perturbation theory analysis seems to imply. In fact, 
if the $\pi^{0}\gamma^{*}\gamma^{*}$ form factor does not change
its sign in the Euclidean range of integration over $k_{i}$, the overall sign is fixed even without knowledge of the form factor. This implies the same positive sign without use of the chiral limit, i.e. 
the same sign for exchanges of heavier pseudoscalars, $J^{PC}=0^{-+}$,  where no large logarithms are present. Moreover, one can verify the same positive sign for exchanges by mesons with $J^{PC}=1^{++},\,2^{-+}$  with an additional assumption about  dominance of one of the form factors. 
Exchanges with  $J^{PC}=0^{++},\,1^{-+},\,2^{++}$ give, however, contributions with a negative sign 
to $a^{\rm HLbL}$ under similar assumptions, but they are much smaller.\\[0.1cm]

\noi
{\sc Next--to--leading terms in the large $N_{c}$ limit}\\[2mm]
Now let us turn to the next--to--leading terms in $1/N_{c}$ expansion. 
Generically these terms are due to two--particle exchanges in the HLbL amplitude, see the diagram in Fig.\,2(b)
with $\pi^{+}\pi^{-}$ substituted by any two meson states. What is specific about the charged
pion loop is its strong chiral enhancement which is not just logarithmic but power--like in this case.
In Eq.\,(\ref{param}) it is reflected in the term $c_{2}\  m_{\rho}^{2}/m_{\pi}^{2}$\,.
The point--like pion loop calculation which gives $a^{\rm HLbL}(\pi\pi)=-4.6\times 10^{-10}$ 
corresponds to $c_{2}=-0.065$. The rather small value of $c_{2}$ can be contrasted 
with the one of the coefficient $c_{1}$ which is not suppressed: $c_{1}\approx 1.7$. 
As we will see the smallness of  $c_{2}$ is related to the fact that chiral perturbation theory does not work in this case.
To see that this is indeed what happens is sufficient to compare the point--like loop result with
the model dependent calculations where form factors are introduced. 
Two known results, $a^{\rm HLbL}(\pi\pi)=-(0.4\pm 0.8)\times 10^{-10}$ \cite{HKS96,HK98} and $a^{\rm HLbL}(\pi\pi)=-(1.9\pm 0.5)\times 10^{-10}$ \cite{BPP96,BPP02},
show a 100\% deviation from the point--like number. It means that the bulk of the contribution does not come from small virtual momenta $k_{i}$ and, therefore, chiral perturbation theory should not be applied. In other words, the term $c_{3}$ in 
Eq.\,(\ref{param}) with no chiral enhancement is comparable with  $c_{2} (m_{\rho}^{2}/m_{\pi}^{2})$.
It means that loops with heavier mesons should also be included.

Breaking of the chiral perturbation theory looks surprising at first sight.
Indeed, the inverse chiral parameter $m_{\rho}^{2}/m_{\pi}^{2}\approx 30$ is much larger
than $N_{c}=3$. What happens is that the leading terms  in the chiral expansion are numerically
suppressed, which makes chiral corrections governed not by $m_{\pi}^{2}/m_{\rho}^{2}$ but 
rather by $\approx 40\, m_{\pi}^{2}/m_{\rho}^{2}$\,. This can be checked analytically in the case of the HVP 
contribution to the muon anomaly. The charged pion loop is also enhanced in this case by a factor
$m_{\rho}^{2}/m_{\pi}^{2}$  but the relative chiral correction due to the pion electromagnetic radius (evaluated with a cutoff at $m_{\rho}^2$ in the $\pi\pi$ spectral function) is 
$\sim 40\, m_{\pi}^{2}/m_{\rho}^{2}\ln(m_{\rho}/2m_{\pi})$. Of course, if the pion mass (together with the muon mass) would be, say, 5 times smaller than in our real world, the charged pion--loop would dominate both in the HVP  and the HLbL contributions to the muon anomalous magnetic moment.

In concluding this Section, we see that the $1/N_{c}$ expansion works reasonably well,
so one can use one--particle exchanges for the HLbL amplitude. On the other hand, chiral enhancement factors are unreliable, so we cannot limit ourselves to the lightest Goldstone--like states, and this is the case both for the leading 
and next--to--leading order in the $1/N_{c}$ expansion.

\vspace*{0.6cm}
\noi
{\bf 3. Short--Distance QCD Constraints.}\\[2mm]
The most recent calculations of $a^{\rm HLbL}$ in the literature~\cite{KN02,HK02,BPP02,MV04} are all compatible with the QCD chiral constraints and large--$N_c$ limit discussed above. They all incorporate the $\pi^0$--exchange contribution modulated by $\pi^0 \gamma^* \gamma^*$  form factors $\cF (k_i^2,k_j^2)$, correctly normalized to the $\pi^{0}\to \gamma\gamma$ decay width. They differ, however, in the shape of the form factors, originating in different assumptions: vector meson dominance (VMD) in a specific form of Hidden Gauge Symmetry (HGS) in Refs.~\cite{HKS96,HK98,HK02}; a different form of VMD in the extended Nambu--Jona-Lasinio model (ENJL) in Ref.~\cite{BPP96,BPP02}; large--$N_{c}$ models in Refs.~\cite{KN02,MV04}; and on whether or not they satisfy the particular operator product expansion (OPE) constraint  discussed in Ref.~\cite{MV04}, upon which we next comment. 

Let us consider a specific kinematic configuration of the virtual photon momenta $k_{1},k_{2},k_{3}$ in the Euclidean domain. In the limit $q=0$ these momenta form a triangle, $k_{1}+k_{2}+k_{3}=0$,
and we consider the configuration where one side of the triangle is much shorter than the others,
$k_{1}^{2}\approx k_{2}^{2} \gg k_{3}^{2}$\,. When $k_{1}^{2}\approx k_{2}^{2}\gg m_{\rho}^{2}$
we can apply the known operator product expansion for the product of two electromagnetic currents
carrying hard moments $k_{1}$ and $k_{2}$,
\be
\lbl{ope}
\int\!\! d^4x_1\!\!\int\!\! d^4x_2  \,{\rm e}^{-ik_1\cdot x_1\! -\!ik_2\cdot x_2} \,
 j_{\nu}(x_1)\ j_{\rho}(x_2) =\frac{2}{\hat k^{2}}\,\epsilon_{\nu\rho\delta\gamma}{\hat k}^{\delta}\!\! \int\!\! d^4z \, {\rm e}^{-ik_{3}\cdot z}\, j_{5}^{\gamma}(z)+{\cal O}\bigg(\frac{1}{{\hat k}^{3}}\bigg)\,.
 \ee
Here $j_{5}^{\gamma}=\sum_{q} Q_{q}^{2}\,\bar q \gamma^{\gamma}\gamma_{5}q$ is the axial current 
where different flavors are weighted by squares of their electric charges and $\hat k=(k_{1}-k_{2})/2\approx k_{1}\approx -k_{2}$\,.  As illustrated in Fig.\,\ref{OPE} this OPE reduces the HLbL amplitude, in 
the special kinematics under consideration, to the AVV triangle amplitude.

\begin{figure}[h]

\begin{center}
\includegraphics[width=0.5\textwidth]{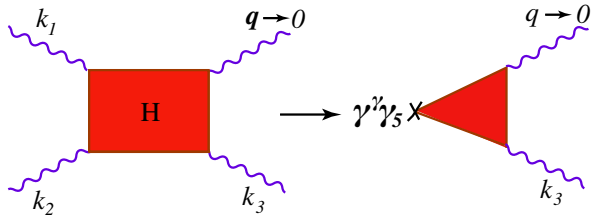}

\vspace*{0.1cm}
\caption{\it\small  OPE relation between the HLbL scattering and the AVV triangle amplitude. }\label{OPE}
\end{center}

\end{figure}

There are a  few things we can learn from the OPE relation in Eq.~\rf{ope}. The first one is 
that the pseudoscalar and pseudovector meson exchanges are dominant at large $k_{1,2}$.
Indeed, only $0^{-}$ and $1^{+}$ states are coupled to the axial current. 
It also provides the asymptotic behavior of form factors at large $k_{1}^{2}\approx k_{2}^{2}$.
In particular, we see that the $\pi^{0}\gamma^{*}\gamma^{*}$ form factor ${\cal F}(k^{2}, k^{2})$
goes as $1/k^{2}$ and similar asymptotics hold for the axial--vector couplings. 
The relation  in Eq.~\rf{ope} does not imply that other mesons, like e.g.  scalars, 
do not contribute to HLbL, it is just that their $\gamma^{*}\gamma^{*}$ form factors 
should fall off faster at large $k^{2}_{1,2}$\,. 

The AVV triangle amplitude consists  of two parts: the anomalous, longitudinal  part and the non--anomalous, transverse one; we consider the chiral limit where $m_{\pi}^{2}\ra 0$.
Because of the absence of both perturbative and non--perturbative corrections to the anomalous AVV triangle graph in the chiral limit, the pion pole description for the isovector  part of the axial current 
works at all values of $k_{3}^{2}$ connecting  regions of soft and hard virtual momenta.
This, in particular, implies the absence of a form factor  $\cF (0,k_3^2)$ in the vertex which contains 
the external magnetic field. At first sight, this conclusion seems somewhat puzzling  because for non--vanishing external 
momentum $q$ the form factor $\cF (q^{2},k_3^2)$ certainly is attributed to the pion exchange.
The answer is provided by the observation that this form factor enters not in the longitudinal anomalous part,
but in the transverse part. It is for this reason that the axial anomaly is not corrected by the form factor.
In the transverse part the form factor shows up together with the massless pion pole in the form
\be
\frac{\cF (q^{2}\!,  k_3^2)-\cF (0,  0)}{(k_{3}+q)^{2}}\,.
\ee
At $q=0$ this combination contains no pion pole at $k_{3}^{2}=0$\,. It means that 
the discussed piece conspires with the pseudovector exchange  to produce 
the transverse result and in this sense becomes part of what could be called 
the pseudovector exchange. It provides the leading short--distance constraint for the pseudovector exchange.
Contrary to the case of the longitudinal component, the transverse, non--anomalous part of the AVV triangle is, however, corrected non--perturbatively~\cite{V03,KPPdeR04}.

Additional constraints on subleading terms in the   $\cF (k_i^2,k_j^2)$ form factor, which were derived in Ref.~\cite{NSVVZ84}, are also taken into account in the calculation quoted in Ref.~\cite{MV04}. 

The large momentum behavior which singles out pseudoscalar and pseudovector exchanges 
is, however, not sufficient to fix {\it per se} a unique model for the evaluation of  $a^{\rm HLbL}$ because the bulk of the integral in Eq.~\rf{aH}
comes from momenta $k_{i}$ of the order of an hadronic scale. However, the faster decreasing of
exchanges other than pseudoscalar and pseudovector ones makes these contributions numerically smaller.
Moreover, the importance of asymmetric momenta configurations with two momenta much larger than the third one was checked in \cite{MV04,BP07} numerically.
This check is related to a question which we next discuss.

There are  other short--distance constraints than those associated with the particular kinematic configuration governed by the AVV triangle. At present, none of the light--by--light hadronic parameterizations made so far in the literature can claim to satisfy fully all the QCD short--distance properties of the HLbL amplitude which is needed for the evaluation of Eq.~\rf{aH}. In fact, within the large--$N_c$ framework, it has been shown \cite{BGLP03} that, in general, for other than two--point functions and two--point functions with soft insertions,  this requires the inclusion of an infinite number of narrow states. However, a numerical dominance of certain momenta configuration could help.
In particular,  in the model of Ref.\,\cite{MV04} with a minimal set of pseudoscalar and pseudovector exchanges,  the corrections due to additional constraints not satisfied in the model turn out to be quite small numerically.
Note that in the frameworks of the ENJL model \cite{BPP96,BPP02} the QCD short--distance constraints 
are accounted for by adding up the quark loop with virtual momenta larger than the cutoff scale of the model.\\

\noi
{\bf 4. Hadronic Model Calculations}\\[2 mm]    
In the previous section we have mentioned a few models used for the calculations of $a^{\rm HLbL}$:
HGS model in \cite{HKS96,HK98,HK02}, ENJL model in \cite{BPP96,BPP02}, the pseudoscalar exchange only in \cite{KN02}, the OPE based 
model of pseudoscalar and pseudovector exchanges in \cite{MV04}. In order to compare  different
results it is convenient to separate the hadronic light--by--light contributions  which are leading in 
the $1/N_c$--expansion from the non-leading ones \cite{deR94}. 

\vspace*{0.3cm} 
\noi
{\sc Contributions which are leading in the $1/N_c$ expansion} \\[2mm]
 Among these contributions, the pseudoscalar meson exchanges which incorporate the $\pi^0$, and to a lesser degree the $\eta$ and $\eta'$ exchanges, are the dominant ones. As discussed above,  there are good QCD theoretical reasons for that. In spite of the different definitions of the pseudoscalar meson exchanges and the associated choices of the  $\cF (k_i^2,k_j^2)$ form factors used in the various model calculations, there is a reasonable agreement among the final results, which we reproduce in Table 1.
\begin{table*}[h]
\caption[RG]{\it Contribution to $a^{\rm HLbL}$ from $\pi^0$, $\eta$ and $\eta'$ exchanges}
\lbl{table1}
\begin{center}
\begin{tabular}{|c|c|} \hline \hline {}&{}\\[-2.5mm] {\bf Result} & {\bf Reference} 
\\[-2.5mm] {}&{}
\\ \hline \hline
{}&{}\\[-2.5mm] 
$(8.5\pm 1.3)\times 10^{-10}$ & \cite{BPP96,BPP02}\\[1mm]
$(8.3\pm 0.6)\times 10^{-10}$ & \cite{HKS96,HK98,HK02}\\[1mm]
$(8.3\pm 1.2)\times 10^{-10}$ & \cite{KN02}\\[1mm]
$(11.4\pm 1.0)\times 10^{-10}$ & \cite{MV04}\\[1mm]
\hline\hline
\end{tabular}
\end{center}
\end{table*} 

\noi
In fact, the agreement is better than this table shows. One should keep in mind 
that in the ENJL model (the first line) the momenta higher than a certain cutoff
are accounted separately  via quark loops while in the OPE based model these momenta
are already included into the result (the last line in the Table 1). Assuming that the bulk of the 
quark loop contribution is associated with the pseudoscalar exchange channel one gets $10.7\times 10^{-10}$ in the ENJL model instead of $8.5\times 10^{-10}$. In the calculations quoted in the two other entries, 
the higher momenta were suppressed by an extra form factor in the soft photon vertex 
and no separate contribution was added to compensate for this.

Closely related to pseudoscalar exchanges is the exchange by the pseudovectors.
Both enter the axial--vector current implying relations between form factors
(see the discussion of the triangle amplitude in the previous section).
 Again, here the estimates in the literature differ by the shape of the form factors used for the $A\gamma^* \gamma^*$ and $A\gamma^* \gamma$ vertex.
Different assumptions on hadronic mixing is another source of uncertainty. Although the contribution from axial--vector exchanges is found to be much smaller than the one from the Goldstone--like exchanges by all the authors, the central values, shown in Table~2, differ quite a lot. 
\begin{table*}[h]
\caption[RG]{\it Contribution to $a^{\rm HLbL}$ from axial-vector exchanges}
\lbl{table2}
\begin{center}
\begin{tabular}{|c|c|} \hline \hline {}&{}\\[-2.5mm]  {\bf Result} & {\bf Reference} 
\\[-2.5mm] {}&{}
\\ \hline \hline
 {}&{}\\[-2.5mm]
$(0.25\pm 0.10)\times 10^{-10}$ & \cite{BPP96,BPP02}\\[1mm]
$(0.17\pm 0.10)\times 10^{-10}$ & \cite{HKS96,HK98,HK02}\\[1mm]
$(2.2\pm 0.5)\times 10^{-10}$ & \cite{MV04}\\[1mm]
\hline\hline
\end{tabular}
\end{center}
\end{table*} 
The authors of Ref.~\cite{MV04} attribute this to the influence of the OPE constraint for the  non--anomalous part of  the AVV triangle amplitude, discussed above.  Further study of the discrepancy 
in this channel is certainly needed.

The scalar exchange contributions have only been taken into account in Refs.~\cite{BPP96,BPP02}. In fact, within the framework of the ENJL model, these contributions are somewhat related to the constituent quark loop contribution. The result is:
 \begin{center}
{\it Contribution to $a^{\rm HLbL}$ from Scalar exchanges}~\cite{BPP96,BPP02}
$$-(0.7\pm0.2)\times 10^{-10}\,.$$
\end{center}
It is much smaller than the contribution from the Goldstone--like exchanges and negative.
In comparison with the pseudovector  exchange,  the magnitude for the scalar  
is  a few times smaller than for the pseudovector in the OPE--based model but a few times  larger 
in HGS and ENJL models.

As we discussed in Section 2 there is some number of other C--even mesonic resonances
in the mass interval 1--2 GeV,  not accounted for in the ENJL model, which could contribute 
to $a^{\rm HLbL}$ comparably to the contribution from scalars. These contributions  are of both signs depending on quantum numbers. At the moment we can only guess about their total effect. Thus, it seems reasonable
to use the scalar exchange result rather as an estimate of error associated with these numerous 
contributions.

\vspace*{0.3cm} 
\noi
{\sc Contributions which are subleading in the $1/N_c$ expansion} \\[2mm]
As we discussed in Section 2 the charge pion loop chirally enhanced as 
$m_{\rho}^{2}/m_{\pi}^{2}$ is a priori the dominant contribution in the subleading  $1/N_c$ order. 
It occurs, however,  that the chiral enhancement does not work and  loops involving other heavier mesons  can compete with the simple pion loop contribution. 

The dressed pion loop results are considerably smaller than the one for
the point--like pion. They are presented in Table 3.
\begin{table*}[h]
\caption[RG]{\it Contribution to $a^{\rm HLbL}$ from a dressed pion loop}
\lbl{table3}
\begin{center}
\begin{tabular}{|c|c|} \hline \hline {}&{}\\[-2.5mm] {\bf Result} & {\bf Reference} 
\\[-2.5mm] {}&{}
\\ \hline \hline
{}&{}\\[-2.5mm] 
$-(0.45\pm 0.85)\times 10^{-10}$ & \cite{HKS96,HK98}\\[1mm]
$-(1.9\pm 0.5)\times 10^{-10}$ & \cite{BPP96,BPP02}\\[1mm]
$(0\pm 1)\times 10^{-10}$ & \cite{MV04}\\[1mm]
\hline\hline
\end{tabular}
\end{center}
\end{table*} 
The last line from Ref.~\cite{MV04} is not the result of a calculation. Strictly speaking it represents  an error estimate of the  meson loop contributions subleading in $1/N_{c}$--expansion.
One can probably increase this error to cover the ENJL result in the second line.

\vspace*{0.3cm}
\noi
{\bf 5. Numerical Conclusions and Outlook}\\[2mm]
What final result can one give at present for the hadronic light--by--light contribution to the muon anomalous magnetic moment? 
It seems to us that, from the above considerations, it is fair to proceed as follows:

\vspace*{0.25cm}
\noi
{\it Contribution to $a^{\rm HLbL}$ from $\pi^0$, $\eta$ and $\eta'$ exchanges}\\[1mm]
Because of the effect of the OPE constraint discussed above, we suggest to take as central value the result of Ref.~\cite{MV04} with, however,  the largest error quoted in Refs.~\cite{BPP96,BPP02}:
\be\lbl{eq:main}
a^{\rm HLbL}(\pi\,,\eta\,,\eta')=(11.4\pm 1.3)\times 10^{-10}\,.
\ee
Let us recall this central value is quite close to the one in the ENJL model when the short--distance quark loop contribution 
is added there.

\vspace*{0.3cm}
\noi
{\it Contribution to $a^{\rm HLbL}$ from pseudovector exchanges}\\[1mm]
The analysis made in Ref.~\cite{MV04} suggests that the errors in the first and second entries of Table~2 are likely to be underestimates. Raising their $\pm 0.10$ errors  to $\pm 1$ puts the three numbers in agreement within one sigma. We suggest then as the best estimate at present
\be\lbl{eq:PV}
a^{\rm HLbL}(\rm{pseudovectors})=(1.5\pm 1)\times 10^{-10}\,.
\ee

\vspace*{0.15cm}
\noi
{\it Contribution to $a^{\rm HLbL}$ from scalar exchanges}\\[1mm]
The ENJL--model should give a good estimate for these contributions. We keep, therefore, the result of Ref.~\cite{BPP96,BPP02} with, however, a larger error which covers the effect of other unaccounted meson exchanges,
\be
a^{\rm HLbL}(\rm{scalars})=-(0.7\pm 0.7)\times 10^{-10}\,.
\ee

\vspace*{0.2cm}
\noi
{\it Contribution to $a^{\rm HLbL}$ from a dressed pion loop}\\[1mm]
Because of the instability of the results for the charged pion loop and unaccounted loops of
other mesons, we suggest using the central value of the ENJL result but wit a larger error:
\be\lbl{eq:dpl}
a^{\rm HLbL}(\pi{\rm -dressed~loop})=-(1.9\pm 1.9)\times 10^{-10}\,.
\ee

From these considerations, adding the errors in quadrature, as well as the small charm contribution in Eq.~\rf{charm}, we get
\be
a^{\rm HLbL}=(10.5\pm 2.6)\times 10^{-10}\,,
\ee
as our final estimate.
 
We wish to emphasize, however, that this is only what we consider to be our best estimate at present. In view of the proposed new $g_{\mu}\!-\!2$ experiment, it would be nice to have more independent calculations in order to make this estimate more robust. More experimental information on the decays $\pi^0 \ra \gamma \gamma^*$, $\pi^0 \ra \gamma^* \gamma^*$ and $\pi^0 \ra e^+ e^-$ (with radiative corrections included) could also  help to confirm the result of  the main contribution in Eq.~\rf{eq:main}. 

More theoretical work is certainly needed for a better understanding of  the other contributions which, although smaller than the one from pseudoscalar exchanges, have nevertheless large uncertainties. 
This refers, in particular, to pseudovector exchanges in Eq.~\rf{eq:PV} but other C-even exchanges 
are also important. Experimental data on radiative decays and two-photon production of  C-even
resonances could be helpful. An evaluation of $1/N_c$--suppresed loop contributions present even a more difficult task. New approaches to the dressed pion loop contribution, in parallel with  
experimental information on  the  vertex $\pi^+\pi^-\gamma^* \gamma^*$, would be very  welcome.
Again, measurement of the two-photon processes  like  $e^+ e^-\to e^+ e^- \pi^+ \pi^-$ 
could give some information on that vertex and help to reduce
the model dependence and therefore the present uncertainty  in Eq.~\rf{eq:dpl}. \\[1mm]  

\noi
{\bf Acknowledgments}\\[2 mm]    
AV is thankful to H. Leutwyler, K. Melnikov and A. Nyffeler  for helpful discussions.
The work of JP and EdeR  has been supported in part by the EU
RTN network FLAVIAnet [Contract No. MRTN-CT-2006-035482]. Work by JP has also been supported
by MICINN, Spain [Grants No. FPA2006-05294 and Consolider-Ingenio
2010 CSD2007-00042 --CPAN--] and by Junta de Andaluc\'{\i}a
[Grants No. P05-FQM 101, P05-FQM 467 and P07-FQM 03048].
The work of AV has been  supported in part by DOE grant DE-FG02-94ER408.

\newpage
\vspace*{1.0cm}

\vfill
 
\end{document}